\documentclass{emulateapj}

\def\gtsim{
\mathrel{\raise.3ex\hbox{$>$}\mkern-14mu\lower0.6ex\hbox{$\sim$}}
}
\def\ltsim{
\mathrel{\raise.3ex\hbox{$<$}\mkern-14mu\lower0.6ex\hbox{$\sim$}}
}
\def\farcs{\hbox{$.\!\!^{\prime\prime}$}}

\def\fs{\hbox{$.\!\!^{\rm s}$}}

\slugcomment{Submitted to the Astrophysical Journal }

\shorttitle{Cas A's Central X-ray Point Source}
\shortauthors{Fesen, Pavlov, \& Sanwal }
\received{2005 June 28}
\begin{document}

\title{Near-Infrared and Optical Limits for the Central \\ 
       X-ray Point Source in the Cassiopeia A Supernova Remnant\altaffilmark{1}}

\altaffiltext{1}{Based on observations with the NASA/ESA Hubble Space Telescope,
obtained at the Space Telescope Science Institute,
which is operated by the Association of Universities for Research in 
Astronomy, Inc.\  under NASA contract No.\ NAS5-26555.
These observations are associated with programs GO-8692 and GO-9798.}

\author{R. A. Fesen\altaffilmark{2}, G. G. Pavlov\altaffilmark{3}  
        \& D. Sanwal\altaffilmark{3} }
\altaffiltext{2}{6127 Wilder Lab, Department of Physics \& Astronomy, Dartmouth College, 
                 Hanover, NH 03755; fesen@snr.dartmouth.edu} 
\altaffiltext{3}{The Pennsylvania State University, 525 Davey Lab., University Park, PA 16802; 
                 pavlov@astro.psu.edu, divas@astro.psu.edu}

\begin{abstract}

We set new near-infrared and optical magnitude limits for the central X-ray
point source (XPS) in the Cassiopeia A supernova remnant based on {\sl HST}
images.  Near-infrared images of the center of Cas~A taken with the NICMOS 2
camera in combination with the F110W and F160W filters ($\sim$ J and H bands)
have magnitude limits $\geq$26.2 and $\geq$24.6, respectively. These images
reveal no sources within a 1\farcs2 radius (corresponding to a 99\% confidence
limit) of the {\sl Chandra} XPS position.  The NICMOS data, taken together with
broadband optical magnitude limits ($R \sim 28$ mag) obtained from a deep STIS
CCD exposure taken with a clear filter (50CCD), indicate that the XPS
luminosities are very low in the optical/NIR bands (e.g., $L_H < 3\times
10^{29}$ erg s$^{-1}$) with no optical, J, or H band counterpart to the XPS
easily detectable by {\sl HST}.  The closest detected object lies 1\farcs8 from
the XPS's nominal coordinates, with magnitudes  $R = 25.7$, $m_{\rm F110W} =
21.9$, and $m_{\rm F160W} = 20.6$, and is a foreground, late-type star as
suggested by Kaplan, Kulkarni, \& Murray.  We discuss the nature of the Cas A
central compact object based upon these near-infrared and optical flux limits.

\end{abstract}


\section{Introduction}

Cassiopeia A (Cas~A) is a $\sim$300 yr old supernova remnant (SNR) containing
high-velocity ejecta exhibiting O and Si-group abundances like those expected
from the supernova of massive star
\citep{ck78,ck79,vdbK83,Tsunemi86,Jansen88,Douvion99,Thor01}.  Because it is
the best known member of the class of remnants from high-mass,
core-collapse supernovae, it has been the target of several investigations
looking for a possible compact stellar remnant near its expansion center
\citep{KvdB76,vdb86,Woan93,Lorimer98}.

None of these searches were successful until first-light images of Cas A taken
in 1999 by the {\sl Chandra} X-ray Observatory revealed a central X-ray point
source (XPS) \citep{Tananbaum99}.  Re-examinations of archival {\sl ROSAT} and
{\sl Einstein} X-ray data \citep{Aschen99,Pavlov99} showed that this X-ray
point source had actually been detected much earlier but was not realized as
such until the {\sl Chandra} higher resolution images clarified 
the spatial structure of the Cas A's central region.

A blackbody fit to the observed {\sl Chandra} Advanced CCD Imaging Spectrometer
(ACIS) spectrum shows a high temperature $kT \approx 0.5$ keV and a small
effective radius $R\approx 0.4$ km, at the remnant's estimated distance of 3.4
kpc \citep{Reed95}, leading to suggestions of a neutron star with hot spots
\citep{Pavlov00}.  Analysis of a recent 1~Ms {\sl Chandra} image of Cas A 
shows no extended pulsar wind nebula around the point source \citep{Hwang04}. 

Although ostensibly the remnant's central compact object (CCO), the nature of
Cas~A's XPS remains uncertain.  The observed X-ray emission could arise either
from a young but radio-quiet neutron star or an accretion disk around a black
hole or neutron star. Soon after its discovery, \citet{Pavlovetal99} and
\citet{Chak01} suggested it might be related to the class of slowly rotating
neutron stars known as Anomalous X-ray Pulsars (AXPs) and Soft Gamma Repeaters
(SGRs). AXPs and SGRs are radio-quiet, exhibit relatively long periods of
several seconds and occasional outbursts, and are thought to have extremely strong
magnetic fields, $\sim 10^{14}$--$10^{15}$ G.  Several AXPs have been
associated with SNRs \citep{Mere02a,Gaensler04} and
a mid-20$^{th}$ century SGR flare from the Cas~A XPS has
been proposed as the source of apparent 24$\mu$ light echo filaments
located $\sim 20'$ north and south of the remnant seen in {\sl
Spitzer} and ground-based K$_{s}$ band images \citep{Krause05}.
However, X-ray period searches on Cas A's XPS have been unsuccessful
so far \citep{Chak01,Mere02b}.

Several deep radio, infrared, and optical searches for central compact object
have been conducted both before and after discovery of the XPS, all without
success. No radio pulsar has been detected at the XPS's position down to 30 mJy
and 1.3 mJy at 327 and 1435 MHz, respectively \citep{McLaughlin01}, and there
is no Infrared Space Observatory ({\em ISO}) detected IR source near the
remnant's center \citep{Lagage96}.  Deep ground-based optical and near-infrared
(NIR) searches have detected only one candidate near the {\sl Chandra} XPS
position ($m_{\rm F675W} =26.7$, $J = 21.4$, $H = 20.5$, $K_s = 20.5$; Kaplan,
Kulkarni, \& Murray 2001).  However, this object appears to be a
foreground Pop II star \citep{Kaplan01}.  Consequently, current optical and NIR
limits (see Table 1) indicate an X-ray to optical flux ratio greater than a few
hundred, thereby excluding some types of accreting binary systems but not
single neutron stars.

Here we report on deep NIR {\sl Hubble Space Telescope} ({\sl HST}) images of
the Cas A XPS region using the Near Infrared Camera and Multi-Object
Spectrograph (NICMOS). We also present a more complete description and analysis
of broadband {\sl HST} optical images obtained with the Space Telescope Imaging
Spectrograph (STIS) \citep{Fes02}. Combined, these data represent the deepest
optical and NIR searches for a counterpart to the Cas~A XPS to date, and we
discuss the implications of these non-detections on the nature of the X-ray
source.

\section{Observations}

\subsection{Near-Infrared Images}
                                                                                                                                          
NICMOS Camera 2 (NIC2; field of view [FOV] of $19\farcs2 \times 19\farcs2$; $0\farcs075$ 
pixel$^{-1}$) was used to image the central region of Cas A during two separate
pointings in 2004 March 3--7 and 2004 June 23--24.  The images were obtained
using the F110W (`NICMOS J') filter ($\lambda_{\rm central}$ = 1.1 $\mu$m,
bandpass = 0.8 -- 1.4 $\mu$m) and F160W (`NICMOS H') filter ($\lambda_{\rm
central}$ = 1.6 $\mu$m, bandpass = 1.4 -- 1.8 $\mu$m).  Both F110W and F160W
images were obtained using a NICMOS sequence of Step=32, Nsamp=25 and a
nine-point, spiral dither pattern, with an exposure time of 576~s per
pointing. A three-orbit, continuous viewing zone (CVZ) observation in March
2004 resulted in a total F160W exposure of 15,550 sec.  Because of a NICMOS
software error, F110W data were obtained on only one of five planned CVZ orbits
in March 2004.  The full five-orbit F110W observations were then repeated in
June 2004.  However, the passages through the South Atlantic Anomaly (SAA)
during the June observations caused a very high cosmic ray background.  We
corrected the SAA affected F110W data using the {\em saa\_clean}\footnote{ {\tt
http://www.stsci.edu/hst/nicmos/tools/post\_SAA\_tools.html}} tool provided by
STScI.  The total exposure in the F110W  band (one CVZ orbit in March and five
CVZ orbits in June) was about 31,000 sec.

Coordinates on the NIC2 images were matched to those of the STIS CCD images
(see \S2.2) by manually matching three stars on each image set.
The data were reduced using IRAF/STSDAS\footnote{IRAF is distributed 
by the National Optical Astronomy
Observatories, which is operated by the Association of Universities for
Research in Astronomy, Inc.\ (AURA) under cooperative agreement with the
National Science Foundation. The Space Telescope Science Data Analysis System
(STSDAS) is distributed by the Space Telescope Science Institute.}  routines.
Measured F110W and F160W fluxes were converted to photometric magnitudes 
on the Vega system using NICMOS photometric keywords and Vegamag 
zeropoints for 77.1 K detector temperatures from current on-line NIC2 
tables\footnote{ {\tt http://www.stsci.edu/hst/nicmos/performance/photometry}}. 

\subsection{Optical Images}

A series of five STIS CCD $52''\times 52''$ images ($0\farcs05$ pixel$^{-1}$)
using the clear ``50CCD'' filter ($\lambda_{\rm center}$ = 585 nm, FWHM = 440
nm) with a total exposure time of 12,402 sec was obtained in January 2001
covering the central region of the Cas A SNR \citep{Fes02}. The data were taken
using a standard STIS CCD dither pattern and reduced using IRAF/STSDAS
reduction programs.

Coordinates of sources detected on the STIS images were measured through the
matching of these sources with those obtained from ground-based images which
were placed on a World Coordinate System (WCS) based on the USNO-A2.0 catalog
\citep{Monet98} in the same manner as described in Thorstensen, Fesen, \& van
den Bergh (2001). Resulting star positions on a $8' \times 8'$ FOV MDM
Observatory 2.4 m R band image  ($2 \times 720$ s exposures) taken in October
1999 were then checked using four Tycho-2 stars detected on the image.  This
resulted in a coordinate registration with the International Coordinate
Reference System (ICRS) to within $0\farcs2$.  Using the ICRS catalog of
reference star positions, we identified three stars on the STIS image that were
then calibrated to the same zero-point of the coordinate grid.  

Figure 1 shows the STIS CCD FOV marked on a 1999 MDM 2.4 m R-band image of Cas
A. The exact positioning of the STIS images was dictated by an attempt to avoid
scattered light problems from relatively bright stars surrounding the X-ray
point source position.  In addition, in order to go as deep as possible, the
data were taken in ``Low-Sky'' mode. This mode minimizes Zodiacal Light which
is the principal background source for wavelengths longer than 3500 \AA.
Low-Sky mode requires the data to be taken at a time during the year when the
Zodiacal Light is no more than 30\% greater than the minimum value for the
Zodiacal Light for the target. 

The STIS 50CCD filter has a very broad effective bandpass, with significant
response from 250 to 1000 nm with a peak near 585 nm.  Consequently, 
STIS 50CCD magnitudes have little direct color information and are sometimes
translated into a standard filter set by quoting the result as a V magnitude.
Knowledge of an object's intrinsic spectrum is required for an accurate
conversion to any standard filter system and various conversion 
techniques have been used \citep{Rejkuba00,Nakamura01,Sollerman02}.
Considering the large extinction toward the remnant (A$_{\rm V} = 5 - 6$ mag;
\citealt{HF96}), which might be even greater toward remnant center
\citep{Troland85}, we have made use of previous R magnitude estimates of field
stars in the STIS CCD FOV to calibrate the STIS data assuming an AB zeropoint of
26.39 mag.

\section{Position of the X-ray Point Source}

A fundamental aspect of searching for an optical or NIR counterpart involves
knowing the precise XPS coordinates along with a positional error radius at a
given confidence level.  Table 2 lists the position of the XPS as cited by
\citet{Tananbaum99}, \citet{Kaplan01}, and \citet{Murray02}.  The optical and
NIR search by \citet{Kaplan01} adopted an XPS position different from that
cited by \citet{Tananbaum99}.  \citet{Kaplan01} found one star (Star A) within
their quoted 90\% confidence level circle (radius = $2\farcs3$). Below, we
attempt to refine the XPS's coordinates using more recent {\sl Chandra} data.  

WCS positions on {\sl Chandra} HRC and ACIS images are believed to be good to
about $\pm 0\farcs6$ (for sources observed close to the aim point) as judged
from comparisons of Tycho-2 and ICRS sources on {\sl Chandra} images (CXO
manual).  Based on 135 target pointings, ACIS-S positions are the most reliable 
(vs.\ ACIS-I, HRC-S, and HRC-I), with a resulting 90\% confidence radius of
$0\farcs55$ (T.\ Aldcroft, priv.\ comm.).  Positional analysis of HRC-S and
HRC-I {\sl Chandra} images have been found to have 90\% confidence radii
similar to that of the ACIS-S, namely $0\farcs65$ and $0\farcs52$, but measured
from fewer pointings (63 and 30, respectively; T.\ Aldcroft, priv.\ comm.).

In Table 2, we list XPS coordinate values derived from several ACIS and HRC
exposures. The HRC positions include small corrections to the normal data
pipeline reduction caused by small aspect offsets (M.\ Juda, priv.\ comm.).
Applying these corrections, the positions of the XPS derived from the three HRC
and two ACIS images all agree within $0\farcs5$.

As an additional check, we have compared our optical coordinate grid directly
with that of the Cas A  {\sl Chandra} observations through an examination of
the recent 1 Ms exposure from the {\sl Chandra} Very Large Program (VLP) ACIS-S
observation of the Cas A remnant \citep{Hwang04} on which a few weak point
sources appear visible around and outside the remnant.  Most of the point
sources on 1 Ms image were close to optical stars, and we manually matched
three of these point sources to stars on our optical reference grid.  Table 3
lists these three reference point sources manually matched, three other optical
sources in the USNO-A2.0 catalog of astrometric stars \citep{Monet98} which
subsequently were found to match X-ray point sources within $1''$ in the
combined 1 Ms {\sl Chandra} ACIS image using our plate solution, plus three
additional sources that matched optical stars (one bright and two faint) on our
reference MDM 2.4 m R-band image. The resulting uncertainty of $0\farcs6$
includes the systematic uncertainty of the USNO-A2.0 catalog \citep{Monet98},
the $0\farcs2$ registration error between the MDM image, and the USNO catalog
\citep{Thor01}, and measurement uncertainty of the XPS on the 1 Ms exposure due
to an asymmetric point spread function.  Our measured position for the XPS
using this revised WCS 1~Ms ACIS-S image is listed near the bottom of Table 2
and is in agreement with the mean of HRC and ACIS derived positions.

Based on this and the  ACIS and HRC positional data, we adopted a XPS position
of $\alpha$(J2000) = 23$^{\rm h}$ 23$^{\rm m}$ $27 \fs 943$, $\delta$(2000) = 58$^{\rm o}$
$48'$ $42 \farcs 51$ with an $1 \sigma$ uncertainty of $0\farcs4$ (see Table
2).  Ironically, this position is in excellent agreement with the initial XPS
coordinates quoted by \citet{Tananbaum99} based only on short (a few ks)
first-light {\sl Chandra} images. 
 
\section{Results and Analysis}

\subsection{Optical and NIR Flux Limits for the XPS} 

The combined STIS 50CCD image of the central region of Cas A roughly centered
on the XPS coordinates is shown in the upper left hand panel of Figure 2.  The
location of the remnant's center of expansion as determined by \citet{Thor01}
is also shown, some $6\farcs6$ to the north.  Two faint stars located $6''$ --
$7''$ west of the XPS coordinates are labeled `1' and `2' in this figure. 

The upper right hand panel of Figure 2 shows a blowup of the STIS image with
95\% confidence circles (radii = $0\farcs9$) centered on the XPS position as
determined from the ACIS and HRC data.  The image shows only one prominent star
near the {\sl Chandra} error circles.  This Star A, discussed by
\citet{Kaplan01}, has a quoted WFPC2 F675W magnitude $m_{\rm F675W} = 26.7\pm
0.2$ and lies outside both 95\% confidence circles.  Based on the star's 
colors, \citet{Kaplan01} argue that it is likely a foreground late M
star, not associated with the XPS.

No other object was detected within or close to the {\sl Chandra} error circle
in the combined STIS 50CCD image. Adopting the \citet{Ryan01} quoted $R = 24.7$
mag for Star 2 located about $7''$ northwest of the X-ray point source's
position (see their Fig.\ 2) and a WFPC2 F675W magnitude of 26.7$\pm$0.2 for
Star A from \citet{Kaplan01}, then
from our estimated STIS 50CCD 3$\sigma$
limit of 28.5 mag for the XPS (Vega magnitude system), we estimate a limit of
$m_{\rm F675W} \geq 28.9$ and $R \geq 27.8$ mag.  A $m_{\rm F675W} -R$
difference of $\simeq$ +1.1 mag is in line with the expected difference for red
objects (e.g., a source with a $B - V = 1.7$ mag; see the WFPC2
Cookbook).  Our derived R magnitudes have uncertainties of around $\pm 0.3$ mag
due to the unknown spectrum of the XPS.  These detection limits also apply to
any optical emission associated with the remnant's center of expansion located
nearly $7''$ north of the XPS \citep{Thor01}.

The NICMOS F110W and F160W images are shown in the two lower panels of Figure
2. The error circles shown are centered on our adopted XPS position (see Table
2). Although we detected some very faint sources in the vicinity of XPS (in
particular Star B 1\farcs9 south of the XPS), we found no sources within a
1\farcs2 radius circle corresponding to a 99\% confidence limit from the
adopted {\sl Chandra} XPS position based on several measurements (see \S3).  We
estimate the $3\sigma$ upper limits on average fluxes of 0.06 and 0.15 $\mu$Jy
in the F110W and F160W bands, respectively, corresponding to F110W and F160W
magnitudes limits of  26.2 and 24.6 (Vega system).  

Optical and NIR magnitudes for stars near the XPS position are given in Table
4. Also listed are our estimated R, F110W, and F160W magnitudes limits for the XPS based
on these STIS and NICMOS images, along with the K$_{\rm s}$ band 
limit from \citet{Kaplan01}.  The NICMOS data, taken together with the deep optical
magnitude limits from the earlier STIS observations ($R \gtrsim 28$ mag)
indicate that there is no optical, J band, or H band counterpart to the XPS
easily detectable by {\sl HST}. 

\subsection{Comparison of NIR, Optical, and X-Ray Fluxes}

The properties of the central compact object in Cas A can be constrained by both
its X-ray emission spectrum and through comparison of X-ray, optical, and NIR
fluxes and luminosities.  
While comparison of uncorrected (absorbed) X-ray, optical, and NIR fluxes
is fairly straightforward, analysis of multiple ACIS observations of Cas A
has shown somewhat different observed XPS's flux values: e.g., from 8 to 9
$\times 10^{-13}$ erg cm$^{-2}$ s$^{-1}$ in the 0.6--6 keV band. 
Because the 0.6--6 keV band is where most of the detected photons are 
thus giving the most accurate characterization of the observed source flux,
the reported scatter is most likely associated with systematic uncertainties for different
observational setups and uncertain pile-up corrections,
which are significant at the observed ACIS-S count rate of 
about 0.35 counts per frame \citep{Teter05}. 
We adopt $F_{\rm X} = 8 \times 10^{-13}$ erg cm$^{-2}$
s$^{-1}$ for the XPS's observed X-ray flux.  The observed (absorbed) magnitudes
limits listed in Table 4 then yield the following flux ratios: $F_{\rm
X}/F_{\rm R} > 25000$, $F_{\rm X}/F_{\rm F110W} > 8400$, $F_{\rm X}/F_{\rm
F160W} > 11000$, $F_{\rm X}/F_{K_s} > 1700$.

To compare extinction corrected (unabsorbed) X-ray and IR-optical luminosities,
knowledge of the amount of interstellar extinction is required.  Optical,
radio, and X-ray measurements indicate the interstellar extinction toward Cas A
appears to be large and non-uniform, with a tendency for increased absorption
going from the northeast to the west and southwest
\citep{Troland85,Keohane96,Will02}. The optical extinction, A$_{\rm V}$, along
the remnant's northern rim based on optical spectra of ejecta and circumstellar
knots ranges from 4.6 to 6.2 mag, with the largest value found farthest off the
northern rim and about halfway toward the remnant center \citep{HF96}.
Adopting A$_{\rm V} = N_{\rm H}/1.8 \times 10^{21}$ cm$^{-2}$ for a typical
gas-to-dust ratio \citep{Bohlin78,PS1995}, an A$_{\rm V}$ value of 6.2 mag
yields an $N_{\rm H,22}\equiv N_{\rm H}/10^{22}\,{\rm cm}^{-2} =1.1$.

However, X-ray and radio measurements suggest that larger column densities are
more common across other sections of the remnant and may be more appropriate
for the remnant's central region.  X-ray measurements of the remnant's overall
emission spectrum suggest $N_{\rm H,22}$ values ranging  from 0.9 to 2.5
\citep{Hughes00,Will02}, corresponding to A$_{\rm V}$ values of 5 -- 14 mag,
with $N_{\rm H,22}$ values around 1.4 -- 1.5
(A$_{\rm V}$ = 7.8 -- 8.3) near the eastern limb
\citep{Hwang02}.
\citet{Troland85} used H\,I and CO
observations to estimate A$_{\rm V}$ values of $4 - 5$ mag for the northern
rim, $5 - 6$ mag for most of the remnant, a value of 7.3 mag near the center,
and as much as 8 mag or more along the western boundary.  

More recent radio observations of the 6 cm H$_{2}$CO transition
\citep{Reynoso02} show clumpy molecular gas toward (and maybe coincident with)
the remnant, again with signs of a significant increase of the H$_2$ column
density toward the western edge of the remnant. \citet{Reynoso02} also find a
H$_{2}$ column density of $5.7 \times 10^{21}$ cm$^{-2}$ for a small molecular
cloud just east of the remnant's expansion center \citep{Thor01}, from which
they estimate $N_{\rm H,22} = 1.3$ (A$_{\rm V}$ = 7.2 mag). The outskirts of
this cloud may extend over to the XPS position. If correct, then this $N_{\rm
H}$ value when added to the absorption of the local arm ($1.3 \times 10^{21}$
cm$^{-2}$; A$_{\rm V}$ = 0.7 mag) suggests a total A$_{\rm V}$ value as high as
8 mag \citep{Reynoso02} toward the XPS.  

Based on the above estimates, we adopt A$_{\rm V}$ = 6 -- 8 mag as a plausible
range for the optical extinction toward the XPS.  For A$_{\rm V}$/E(B--V) =
3.1, this converts to A$_{\rm R} = 4.5 - 6.0$, A$_{\rm J} = 1.7 - 2.2$, A$_{\rm
H} = 1.0 - 1.4$, and A$_{\rm K} = 0.6 - 0.9$.  Assuming for instance $A_{\rm
V} = 7$, we obtain the following upper limits on NIR-optical luminosities:
$L_{\rm opt/NIR} < 5.6$, 0.8, 0.3, and 1.3 $\times 10^{30}$ erg s$^{-1}$ for
the R, F110W, F160W, and K$_{\rm s}$ bands, respectively (for $d=3.4$ kpc). 

$N_{\rm H}$ values can be estimated directly from
fitting the X-ray spectra, but the results are very sensitive to
the choice of spectral model.
Fits with
blackbody models yield lowest column densities, $N_{\rm H,22} =0.6$ -- 1.2
in different data sets, and lowest unabsorbed fluxes,
$F_{\rm X}^{\rm unabs} = (1.2$ -- $1.4)\times 10^{-12}$ erg cm$^{-2}$ s$^{-1}$
in the 0.6 -- 6 keV band.
Power-law fits yield highest values for both the column density
and the unabsorbed flux: $N_{\rm H,22} =1.8$ -- 2.6,
$F_{\rm X}^{\rm unabs} = (8$ --$12)\times 10^{-12}$ erg cm$^{-2}$ s$^{-1}$
in the same band.
Fits with more complicated models (neutron star atmospheres,
blackbody plus power law, etc) yield intermediate values, typically
$N_{\rm H,22} = 1.0$ -- 1.2 and $F_{\rm X}^{\rm unabs} =
(1.2$ -- $1.6)\times 10^{-12}$ erg cm$^{-2}$ s$^{-1}$,
corresponding to the isotropic 0.6--6 keV luminosity
$L_X=4\pi d^2 F_{\rm X}^{\rm unabs} = (1.7$ -- $2.2)\times 10^{33}$
erg s$^{-1}$. 

Unfortunately, it is currently unknown which of the models is more appropriate
for the XPS since the true shape of its X-ray spectrum is not completely
certain because of CCD pile-up corrections.  However, the spectrum is likely to
be either thermal or at least has a thermal component which favor intermediate
$N_{\rm H}$ values, $N_{\rm H,22} = 1.1$ -- 1.4, corresponding to A$_{\rm V}$
values of 6 -- 8 (for a typical gas-to-dust ratio) and entirely consistent with
the optically derived absorption estimates discussed above.

Therefore, assuming A$_{\rm V}=7$ and $L_X = 2\times 10^{33}$ erg s$^{-1}$
for the 0.6 -- 6 keV energy range, we obtain the following ratios of
X-ray to optical/NIR unabsorbed luminosities:
$L_X/L_{\rm R} > 360$, $L_X/L_{\rm F110W} > 2500$, $L_X/L_{\rm F160W} > 6700$,
and $L_X/L_{\rm K_{s}} > 1600$.

\section{The Nature of the Cas A XPS}

Analyzes made soon after the discovery of the Cas A XPS indicated that it was
unlikely to be an active pulsar. There was no detected radio pulsation or $\gamma$-ray
emission, no X-ray or radio plerion, and the observed X-ray spectrum appeared
too steep for an ordinary pulsar 
\citep{Pavlov00}. We will show below that the deep
optical-NIR upper limits reported above support this conclusion. 

Optical and X-ray emission of an active pulsar is characterized by a nonthermal
(presumably synchrotron) component emitted from the pulsar's magnetosphere
which dominates at harder X-rays and longer optical-NIR wavelengths while the
main contribution from the thermal component is in the UV and soft X-rays.  An
example of a pulsar's NIR to X-ray spectrum is shown in the middle panel of
Figure 3 for PSR B0656+14 \citep{Pavlov02b,Kargaltsev05}, one of few pulsars
showing thermal emission in both soft X-rays and the UV.  For this pulsar, the
nonthermal component, which dominates at $E\gtsim 2$ keV and $\lambda >
2000$~\AA,  has about the same spectral slope in the hard X-rays as in the
NIR-optical (photon index $\Gamma\approx 1.5$), with ratios of the nonthermal
$0.6-6$ keV  X-ray luminosity to NIR luminosities of $L_{X}/L_{\rm
F110W}\approx 200$ and $L_{X}/L_{\rm F160W}\approx 400$.  In other young
pulsars such as the Crab pulsar, NIR and optical fluxes may lie below the
continuation of the (steeper) X-ray nonthermal spectrum, but the ratios of the
nonthermal X-ray-to-optical luminosities are almost the same, at a level of
about a few hundred \citep{ZavlinPavlov04}.

The Cas~A XPS 
could either have a pure thermal spectrum, which
would support the assertion that it is not an active pulsar, or it may have a
nonthermal power-law component dominating above 4--5 keV with a slope $\Gamma
\approx 2$ and a 0.6--6 keV unabsorbed flux $\sim 2\times 10^{-13}$ erg
cm$^{-2}$ s$^{-1}$ corresponding to $L_{X}\sim 3\times 10^{32}$ erg s$^{-1}$
(Teter et al.\ 2005).  However, an extension of this component into the
optical-NIR range lies well above our upper limits (see Fig.\ 3).  Moreover,
the lower limits on the ratios of nonthermal $0.6-6$ keV  X-ray and NIR/optical
luminosities (e.g., $L_{X}/L_{\rm F110W} \gtsim 370$, $L_{X}/L_{\rm
F160W}\gtsim1000$) are somewhat higher than observed in active pulsars.

If the Cas~A XPS is a neutron star, the lack of manifestations of pulsar activity can
be used to constrain its parameters. Assuming the characteristic age
of a putative pulsar, $\tau =P/2\dot{P}$, is close to its actual age,
$\approx 1\times 10^{10}$ s for Cas A, we obtain $P=(2\pi^2 I/\tau
\dot{E})^{1/2} \approx 1.4\, \dot{E}_{36}^{-1/2}$ s, $\dot{P} \approx 6.8\times
10^{-11} \dot{E}_{36}^{-1/2}$, and $B_p =6.4\times 10^{19} (P\dot{P})^{1/2}
= 6.2\times 10^{14} \dot{E}_{36}^{-1/2}$ G, where $I\simeq 10^{45}$ g cm$^2$ is the
moment of inertia, $\dot{E} = 10^{36}\dot{E}_{36}$ erg s$^{-1}$ is the pulsar's
spindown energy loss rate, and $B_p$ is a conventional estimate on the magnetic
field at the neutron star's magnetic pole.

Based on the observed correlation between the pulsar optical luminosity 
and spin-down energy loss rate (e.g., \citealt{Zharikov04}),
our NIR-optical limits on the XPS's radiation imply a conservative
upper limit $\dot{E} < 10^{37}$ erg s$^{-1}$, 
similar to the limit that follows from the lack of the pulsar wind nebula
\citep{Gotthelf04b}. Such a limit translates into $P>0.4$ s,
$\dot{P} > 2\times 10^{-11}$, and $B_p > 2\times 10^{14}$ G.  This suggests that
the XPS might be a slowly rotating neutron star with a very high magnetic
field.  (Note: The field could be lower than the above estimate if the XPS was born as
slow rotator.)

Fits of the XPS's X-ray spectrum with a blackbody model show that the
emitting area, $\sim 1$ km$^2$, is too small and the temperature, $\approx
6\times 10^6$ K, too high for a $\sim$ 300 yr old neutron star (Pavlov et al.\
2000).  The small size of the emitting area initially hinted that it might be a
strongly magnetized neutron star accreting matter, possibly from a secondary
companion in a binary, albeit with a very low accretion rate, $\dot M\sim 
10^{13}$ g s$^{-1}$.  However, the {\sl HST} optical and NIR observations
described above now place tight constraints on companion's absolute magnitudes.
Adopting $D=3.4$ kpc and A$_{\rm V}$ = 7 mag, the STIS and NICMOS magnitude
limits imply M$_{\rm R}$ $\geq$10, and M$_{F110W}\geq 11.5$, M$_{F160W} \geq
10.7$, fainter than any main-sequence star, let alone secondary components of
known X-ray binaries \citep{vanparadijs95}.
The same argument virtually excludes accretion onto a
black hole in a binary. 

The observed X-ray luminosity of the XPS is also too high to be explained
via accretion from the interstellar medium. At the $\simeq$ 300 km s$^{-1}$
velocity derived from the displacement of the XPS from the SNR's expansion
center \citep{Thor01} and a $\sim$ 300 yr age for the SNR, an
unrealistically high ISM density around $10^5$ cm$^{-3}$ is needed to
obtain the required accretion rate \citep{Pavlov00}.

On the other hand, it is possible that accretion onto the central compact
object might occur, not from a secondary companion or from the ISM, but from a
hypothesized ``fossil accretion disk'' consisting of debris left over from the
supernova explosion (e.g., \citealt{Paradijs95};
\citealt{Eksi05}, and references therein). Formation of an ejecta
accretion disk due to fallback material around the neutron star sometime after
the supernova explosion could give rise to 
NIR emission due to 
X-ray irradiation and/or viscous energy dissipation
\citep{Chat00,Perna00,Eksi03}. Such a disk might be 
identifiable by exhibiting NIR and optical colors substantially
different from stellar ones.  
However, if the XPS's X-ray luminosity is due to accretion from the disk
onto the neutron star surface,
the corresponding accretion rate, $\dot{M}\sim 10^{13}$ g s$^{-1}$, implies
a very low disk mass, $M_d(t) \sim (16/3)\dot{M} t \sim 3\times 10^{-10} M_\odot$
at the present age $t\sim 10^{10}$ s
in the disk model considered by Perna et al.\ (2000). Such a small disk would
hardly be detectable even in very deep NIR observations.

A strong argument against any sort of accretion as being the chief
source of the XPS's X-ray emission
is the lack of noticeable variability.
\citet{Teter04} have analyzed {\sl Chandra}
observations of the Cas A XPS spanning a range of 4.5 years and found no 
statistically significant flux
changes, usually observed in accreting X-ray sources.
Consequently, accretion, regardless of the source of material, would seem to be
an unlikely source of the observed X-ray emission.

It has been suggested that the Cas A XPS may be simply a more youthful and less
luminous example of the isolated neutron star subclasses known as Anomalous
X-ray Pulsars (AXPs) and Soft Gamma-ray Repeaters (SGRs)
\citep{Chak01,Mere02b,Pavlov02a,Pavlov04,Roth03}.  Occasional bursts from AXPs
and SGRs as well as their spin-down rates have been successfully explained by a
magnetar model \citep{DT92,Thompson96}, in which a neutron star has a surface
magnetic field of $10^{14}-10^{15}$ G.  Except for a lack of pulsations, the
general properties of the Cas A XPS and other CCOs in fairly young SNRs are not
all that dissimilar from AXPs and SGRs.  The fact that there are at least two
cases of AXPs associated with SNRs \citep{Gaensler04} might indicate that CCOs
and AXPs are  related.
 
AXPs exhibit X-ray spectra similar to those observed for CCOs, including the
Cas~A XPS (Pavlov et al.\ 2002a,2004), and show X-ray pulsations with periods
in the range of 5--12 s.  So far, however, no pulsations have been found for
the Cas~A XPS.  The Murray et al.\ (2001) claim for a 12.5 ms period from the
{\sl Chandra} HRC observations was based on a low significance (2.5$\sigma$)
signal, and it was not confirmed in a follow-up HRC observation
\citep{Ransom02}.  Moreover, a neutron star with a period as short as this
would be an active pulsar even if it had a very low magnetic field,
$\sim3\times 10^8$ G, and it would be hard to explain the origin of this
periodicity if the field is even lower than that.

While at least four of the seven firmly established
AXPs have now been detected in the optical/NIR, no CCO in a young SNR
has yet been seen. Table 5 lists the observed optical (R band) and NIR (JHK$_{s}$)
magnitudes or limits for the Cas~A CCO along with five other SNRs with 
radio-quiet, X-ray emitting CCOs. Remnants with CCOs similar to that of Cas A are
Puppis A (G260.4$-$3.4), the {\sl ROSAT} discovered remnant G266.2$-$1.2 
(sometimes called `Vela Junior' due to its location near the Vela SNR),
and G347.3$-$0.5.
The CCOs in G296.5+10 and Kes 79, for which short periods have been 
recently detected \citep{Zavlin00,Zavlin04,Gotthelf05},
possibly belong to a separate subclass. 
Table 5 also lists the six Galactic AXPs which have been 
observed in the optical and/or NIR, plus one AXP found in the SMC
(CXOU~J0100$-$7211).
For these six CCOs and seven AXPs, Table 6 lists observed pulsation periods 
(where 
detected), estimated ages (either the associated remnant age or spin-down age),
estimated distance, $N_{\rm H}$ column density, observed (absorbed) 
0.5--10 keV, H, and K$_{s}$ band fluxes, and estimated X-ray (2--10 keV),
H, and K$_{s}$ luminosities\footnote{We chose the H and K$_{s}$ NIR bands
and the 2--10 keV X-ray band as they are less sensitive to the poorly
known interstellar extinction.}.   

As shown in Tables 5 and 6, our new J and H magnitudes limits for a Cas A XPS
counterpart are about 2 magnitudes  deeper than the deepest previous CCO or AXP
searches.  Our NICMOS J and H band magnitudes limits for the XPS are about 4
magnitudes deeper than previous searches for a Cas~A XPS counterpart (see Table
1) and about 2 magnitudes deeper than currently possible from 8 -- 10 m
ground-based telescopes.  However, a K band limit of $\sim 22 - 23$ mag, about $1 -
2$ mag deeper than the 21.2 mag
 limit set by \citet{Kaplan01}, is feasible
using 8 -- 10 m telescopes and exposures of several hours.

In terms of extinction-corrected luminosities, 
we reached the deepest NIR luminosity limit in the H band.
Our limit of $L_{\rm H}\leq 0.3 \times 10^{30}$ erg s$^{-1}$ is about an order
of magnitude
lower than the 
luminosities (or luminosity limits) of the AXPs observed in the H band.
However, the X-ray luminosity of
the Cas A XPS
for the column density insensitive 2--10
keV energy range is also well below those seen for most AXPs
yet is not unlike other CCOs in relatively young remnants.

The origin of the NIR/optical emission in AXPs is unknown.
Although some hypotheses have been suggested in the framework
of the magnetar model \citep{Thompson02,Eichler02},
they do not give
quantitative predictions for IR or optical emission levels.
However, the recently observed correlations between the X-ray and
NIR luminosities in some variable AXPs (most notably in 1E~2259+586;
see Tam et al.\ 2004) suggest that the NIR radiation and X-ray radiation may be 
somehow related, and the NIR luminosity is a certain fraction,
$\sim 10^{-4}-10^{-3}$, of the
X-ray luminosity.
On the other hand, \citet{Durant05a}
did not find such a correlation for the AXP 1E~1048$-$5937.

For the Cas A XPS, the upper limit $L_{\rm Ks}/L_{\rm X} \leq 1.7\times 10^{-3}$
is still above the values of this ratio for the K$_s$ band detected AXPs.
The H band limit, $L_{\rm H}/L_{\rm X} \leq
4\times 10^{-4}$, is a factor of 3 lower than the $L_{\rm H}/L_{\rm X}$ 
value for J1708$-$4009, but the identification of the NIR counterpart
of this AXP remains uncertain \citep{Durant05a}. Nearly simultaneous X-ray and NIR
observations of the variable XTE~J1810$-$197 \citep{Rea04}
showed $L_{\rm H}/L_{\rm X}\approx 0.5\times 10^{-4}$, a factor of 8
lower than for the XPS upper limit. For 1E~1048$-$5937, another X-ray and NIR 
variable AXP, this ratio is apparently variable.
Using the H band flux
from the Magellan observation of 2001 March 24 \citep{Wang_Chak02b} 
and X-ray flux from the {\sl XMM-Newton} observation of 2000 December 28
\citep{Mere04}, we obtain  $L_{\rm H}/L_{\rm X}\approx 1\times 10^{-3}$,
while only an upper limit of $\ltsim 5\times 10^{-4}$ can be inferred
from nearly simultaneous VLT and {\sl XMM-Newton} observations of
2004 June \citep{Durant05a,Mere04}. 

Thus, although it is currently unclear how universal 
the NIR-to-X-ray luminosity ratio is for AXPs, our deep $L_{\rm H}/L_{\rm X}$
limit is comparable with the luminosity ratios observed in AXPs. Perhaps more
conclusive results would be obtained from the comparison of
the $L_{\rm K}/L_{\rm X}$ ratios after much deeper K band observations of
the Cas A XPS are carried out. If such future observations detect
the XPS, and the NIR-to-X-ray luminosity ratio turns out to be in the
same range as for AXPs, it would be a strong evidence that these
objects are indeed related.

A direct comparison of the 
spectral energy distributions for the Cas A's XPS,
the AXP 1E~2259$+586$ in the SNR CTB 109,
and PSR B0656$+14$ 
is shown in Figure 3
along with the measured optical and NIR flux detections or limits.
We see that 
both the XPS and the AXP have similar
thermal components of the blackbody$+$power-law X-ray spectral models
(but the AXP's thermal component is much brighter because of a larger
emitting area), both being hotter than the two thermal components
of the pulsar's spectrum.  The extrapolations of the
XPS's and AXP's thermal components into the
optical domain are well below the observed fluxes
or flux upper limits, while such an extrapolation matches the 
observed UV spectrum of the pulsar.
The extrapolation of the
pulsar's  X-ray nonthermal (power-law) component 
approximately matches the observed NIR-optical spectrum, contrary to the
much softer AXP's power-law component. The slope (and even the
presence) of such a component in the XPS's spectrum is very
uncertain (see \S4); if present, its NIR-optical extrapolation
also lies above the observed upper limits.
Obviously, the broadband (NIR through X-rays) spectrum of 1E~2259+586 (and other
AXPs) cannot be described by a simple model. Moreover, the discrepancy
between the X-ray and NIR spectra suggests that even the generally
adopted blackbody$+$power-law model for the X-ray spectrum may not be
adequate. The nature of the Cas A's XPS broadband spectrum is even
less clear, given the lack of detections in the NIR-optical and the
uncertainty of its hard X-ray tail caused by the CCD pile-up.
What is clear, however, is that the data currently available cannot rule out the hypothesis 
that CCOs and AXPs, being different from the commonly known radio and $\gamma$-ray pulsars, are related. 

Although CCOs, including the Cas A's XPS, are 
the stellar remnants of core-collapse supernovae, the
mass range and possibly binarity of the SN progenitor leading to their
formation is unclear.  The presence of CCOs in the high mass, core-collapse
SNRs of Puppis A and Cas A with similar properties is intriguing given both
remnants' relatively young ages (cf.\ Table 6) and similar ejecta kinematics
and abundances. Both remnants show high-velocity, oxygen-rich ejecta and have
considerable amounts of surrounding CSM material rich in nitrogen suggestive of
a high mass loss episode just prior to explosion like that from a fairly high
mass, WR star progenitor \citep{Fes87}. 

This leads to the question of why don't other young, high mass, O-rich
core-collapse SNRs also possess similar CCOs?  The 1--2 kyr old, O-rich SMC
remnant 1E~0102$-$72 shows no CCO, but this may simply be a problem of
detection difficulty due to its greater distance.  Similarly, for the recently
discovered but much older O-rich SMC remnant, 0103$-$72.6 \citep{Park03}, the
presence of considerable X-ray emission near the remnant center might prevent a
clear CCO detection, much like what happened in the case of Cas A before {\sl
Chandra}'s improved resolution became available.  However, the relatively young
oxygen-rich Galactic and LMC remnants, G292.0+1.8 and 0540$-$69.3, do show
bright, rapidly spinning pulsars ($P=135$ ms and 50 ms, respectively;
\citealt{Camilo02,Seward84,Manchester93}) and both have extensive pulsar wind
nebulae.  Currently, we do not understand which properties of the apparently
similar progenitors predetermine the nature of the compact remnant, a CCO, or an
active pulsar of the SN explosion.

In the end, a key element in our understanding of the nature of the Cas~A CCO,
the Puppis A CCO, and similar objects will be determining their periodicity
as well as an accurate measurement of their X-ray spectra.
While the data frame time employed for the recent 1 Ms {\sl Chandra} 
observation of
Cas~A may not be useful in investigating periods $\lesssim$ 10~s
and obtaining a clean source spectrum,
specially designed future X-ray observations of Cas~A may help to
find a shorter period and obtain an accurate X-ray spectrum. In the NIR-optical
range, much deeper K band observations are possible.
Such X-ray and NIR observations of Cas~A may finally
resolve the nature of its XPS and thus other compact stellar remnants as well. 

\acknowledgments

We thank M. Hammell for help with the optical coordinate matching, P.
Plucinsky, T. Aldcroft, and M. Juda for assistance and guidance with measuring
the XPS position on {\sl Chandra} ACIS and HRC images, and Al Schultz
for help with reduction and analysis of the NICMOS data.  Support for this work
was provided by NASA through grant numbers GO-8692 and GO-9798 from the Space
Telescope Science Institute, which is operated by AURA, Inc., under NASA
contract NAS5-26555. The work of G.G.P. and D.S. was also partially supported
by NASA grant NAG5-10865.

\clearpage

\begin{deluxetable}{llccccl}
\scriptsize
\tablecaption{Optical and Near-Infrared Magnitude Limits for the Cas A XPS }
\tablewidth{0pt}
\tablehead{
\colhead{Reference} & \colhead{R}  & \colhead{F675W} &  \colhead{J}  
   &  \colhead{H}  &  \colhead{K}
}
\startdata
van den Bergh \& Pritchet (1986) ~ ~  & $\geq$24.8 & \nodata    & \nodata    & \nodata   &  \nodata \\
Ryan et al.\ (2001)   & $\geq$26.3 & \nodata    & \nodata    & \nodata   &  \nodata \\ 
Kaplan et al.\ (2001) & $\geq$25.0  & $\geq$27.3 & $\geq$22.5 & $\geq$20  & $\geq$
21.2 \\
Fesen et al.\ (2002) & $\geq$27.8\tablenotemark{a}  & $\geq$28.9\tablenotemark{a}  & \nodata    & \nodata   & \nodata  \\
This work  & \nodata & \nodata & $\geq$26.2\tablenotemark{b} & $\geq$24.6\tablenotemark{c} & \nodata \\
Ground-based NIR limits\tablenotemark{d} & \nodata & \nodata & $\sim$23.6 & $\sim$22.8 & $\sim$22.3 \\ 
\enddata
\tablenotetext{a}{Estimated value from the 50CCD STIS image.}
\tablenotetext{b}{F110W magnitude limit.}
\tablenotetext{c}{F160W magnitude limit.} 
\tablenotetext{d}{Magnitude limits that could be reached in a one hour
exposure (without adaptive optics) using Subaru's IRCS in 0\farcs5 seeing.}
\end{deluxetable}

\begin{deluxetable}{llll}
\scriptsize
\tablecaption{Coordinates of the Cas A X-Ray Point Source (XPS) }
\tablewidth{0pt}
\tablehead{
\colhead{Reference} & \colhead{Source}  & \colhead{$\alpha$(J2000)} & \colhead{$\delta$(J2000)} } 
\startdata
Tananbaum (1999)      & ACIS-S ObsID 0214    & $23^{\rm h} ~ 23^{\rm m} ~ 27 \fs 940$ $\pm 0 \fs 30$ &
          $58^{\circ} ~ 48^{'} ~ 42 \farcs 40$  $\pm 2 \farcs 5$ \\ 
Kaplan et al.\ (2001)
          & ACIS + HRC    & $23^{\rm h} ~ 23^{\rm m} ~ 27 \fs 857$ $\pm 0 \fs 13$ &
          $58^{\circ} ~ 48^{'} ~ 42 \farcs 77$  $\pm 1 \farcs 0$  \\
Murray et al.\ (2002)         & HRC-S ObsID 1857   & $23^{\rm h} ~ 23^{\rm m} ~ 27 \fs 920$ $\pm 0 \fs 13$ &
          $58^{\circ} ~ 48^{'} ~ 42 \farcs 55$  $\pm 1 \farcs 0$  \\
This work                             & HRC-S  ObsID 1038      & $23^{\rm h} ~ 23^{\rm m} ~ 27 \fs 957$ $\pm 0 \fs 08$ &
          $58^{\circ} ~ 48^{'} ~ 42 \farcs 53$  $\pm 0 \farcs 6$  \\
 ~ " ~ ~ "                             & HRC-I~ ObsID 1505    & $23^{\rm h} ~ 23^{\rm m} ~ 27 \fs 932$ $\pm 0 \fs 08$ &
          $58^{\circ} ~ 48^{'} ~ 42 \farcs 57$  $\pm 0 \farcs 6$  \\
 ~ " ~ ~ "                             & HRC-S  ObsID 1857      & $23^{\rm h} ~ 23^{\rm m} ~ 27 \fs 961$ $\pm 0 \fs 08$ &
          $58^{\circ} ~ 48^{'} ~ 42 \farcs 62$  $\pm 0 \farcs 6$  \\
 ~ " ~ ~ "                            & ACIS-S ObsID 5196    & $23^{\rm h} ~ 23^{\rm m} ~ 27 \fs 921$ $\pm 0 \fs 08$ &
          $58^{\circ} ~ 48^{'} ~ 42 \farcs 46$  $\pm 0 \farcs 6$  \\
 ~ " ~ ~ "                            & ACIS-S ObsID 5319   & $23^{\rm h} ~ 23^{\rm m} ~ 27 \fs 913$ $\pm 0 \fs 08$ &
          $58^{\circ} ~ 48^{'} ~ 42 \farcs 34$  $\pm 0 \farcs 6$  \\
 ~ " ~ ~ "                              & ACIS-S VLP + Optical & $23^{\rm h} ~ 23^{\rm m} ~ 27 \fs 945$ $\pm 0 \fs 08$ &
          $58^{\circ} ~ 48^{'} ~ 42 \farcs 45$  $\pm 0 \farcs 6$  \\
\hline
Adopted XPS position       &  & $23^{\rm h} ~ 23^{\rm m} ~ 27 \fs 943$ $\pm 0 \fs 05$ &
          $58^{\circ} ~ 48^{'} ~ 42 \farcs 51$  $\pm 0 \farcs 4$  \\
\enddata
\end{deluxetable}

\begin{deluxetable}{lllcc}
\tablecaption{Point sources on 1 Ms VLP  {\sl Chandra} ACIS-S image of Cas A}
\tablewidth{0pt}
\tablehead{
\colhead{Star ID} & \colhead{$\alpha$(J2000)} & \colhead{$\delta$(J2000)} & \colhead{R mag } &  \colhead{B mag } }
\startdata
\underline{Manually matched USNO-2.0 stars}   &  &  &  & \\
USNO 2.0 U1425-15020099 & $23^{\rm h} ~ 23^{\rm m} ~ 26 \fs 33$ & $58^{\circ} ~ 53^{'} ~ 10 \farcs 1$ & 18.1 & 20.2 \\
USNO 2.0 U1425-15018830 & $23^{\rm h} ~ 23^{\rm m} ~ 22 \fs 78$ & $58^{\circ} ~ 52^{'} ~ 49 \farcs 4$ & 11.7 & 12.5 \\
USNO 2.0 U1425-15027102 & $23^{\rm h} ~ 23^{\rm m} ~ 45 \fs 81$ & $58^{\circ} ~ 50^{'} ~ 48 \farcs 7$ & 12.9 & 14.3 \\
\underline{Sources matching USNO-2.0 catalog}   &  &  &  & \\
USNO 2.0 U1425-15007635 & $23^{\rm h} ~ 22^{\rm m} ~ 51 \fs 55$ & $58^{\circ} ~ 50^{'} ~ 19 \farcs 6$ & 14.8 & 16.5 \\
USNO 2.0 U1425-15012096 & $23^{\rm h} ~ 23^{\rm m} ~ 03 \fs 84$ & $58^{\circ} ~ 49^{'} ~ 18 \farcs 2$ & 14.9 & 16.3 \\
USNO 2.0 U1425-15012463 & $23^{\rm h} ~ 23^{\rm m} ~ 04 \fs 82$ & $58^{\circ} ~ 48^{'} ~ 00 \farcs 1$ & 14.2 & 15.3 \\
\underline{Sources matching non-USNO stars}   &  &  &  & \\
MDM ID 1093 & $23^{\rm h} ~ 23^{\rm m} ~ 43 \fs 16$ & $58^{\circ} ~ 52^{'} ~ 24 \farcs 7$ & 12.8 & \nodata \\
MDM ID 1070 & $23^{\rm h} ~ 23^{\rm m} ~ 41 \fs 76$ & $58^{\circ} ~ 52^{'} ~ 27 \farcs 3$ & 19.5 & \nodata \\
MDM ID 1315 & $23^{\rm h} ~ 23^{\rm m} ~ 57 \fs 17$ & $58^{\circ} ~ 48^{'} ~ 12 \farcs 1$ & 21.2 & \nodata \\
\enddata
\end{deluxetable}

\begin{deluxetable}{clllccccc}
\tabletypesize{\scriptsize}
\tablecaption{Optical and Near-IR Sources Near the Cas A X-Ray Point Source (XPS)}
\tablewidth{0pt}
\tablehead{
\colhead{Object} & \colhead{$\alpha$(J2000)} & \colhead{$\delta$(J2000)} &
\colhead{$\Delta r$} &
\colhead{STIS\tablenotemark{a}}
&  \colhead{R } 
& \colhead{F110W}  & \colhead{F160W}  & \colhead{K$_{s}$ }  }
\startdata
XPS       & $23^{\rm h}  23^{\rm m}  27 \fs 943 \pm 0 \fs 05$ &
          $58^{\circ}  48^{'}  42\farcs 51  \pm 0\farcs 4$ &  $0 \farcs0$  &
          $\geq$ 28.5 & $\geq$ 27.8 & $\geq$ 26.2 & $\geq$ 24.6 & $\geq$ 21.2\tablenotemark{b} \\
          & &                                                     & & & & \\
Star A    & $23^{\rm h}  23^{\rm m}  27 \fs 752 \pm 0 \fs 012$ & 
            $58^{\circ}  48^{'}  41\farcs 51 \pm 0 \farcs 1$ &  $1 \farcs8$ &
            26.4$\pm0.1$ & 25.7$\pm0.2$\tablenotemark{c}
           & 21.9$\pm 0.1$ & 20.6$\pm 0.1 $ & 20.5$\pm 0.3$ \\

Star B    & $23^{\rm h} 23^{\rm m} 27\fs 938 \pm 0\fs 025$ &
          $58^{\circ} 48^{'} 40\farcs 59 \pm 0\farcs 2$ & $1\farcs9$ &
          $\geq28.5$ & $\geq27.8$ 
         & 24.7$\pm 0.2$ & 23.0$\pm 0.2$ & \nodata \\

Star C    & $23^{\rm h} 23^{\rm m} 28\fs 075 \pm 0\fs 020$ &
          $58^{\circ} 48^{'}  45\farcs 92 \pm 0\farcs 1$ &  $3\farcs6$ &
          $\geq28.5$  & $\geq27.8$ 
        & 25.6$\pm 0.2$ & 23.7$\pm 0.2$ & \nodata\\
          & &                                                    &  & & & & \\
Star 1    & $23^{\rm h} 23^{\rm m} 27\fs 170 \pm 0\fs 012$ &
          $58^{\circ} 48^{'} 40\farcs 25 \pm 0\farcs 1$ & $6\farcs5$ &
          23.6$\pm0.1$ & 22.9$\pm0.3$ 
         & 19.1$\pm 0.1$ & 17.4$\pm 0.1$ & 17.5$\pm0.2\tablenotemark{d}$ \\

Star 2  & $23^{\rm h} 23^{\rm m} 27\fs 215 \pm 0\fs 012$ &
          $58^{\circ} 48^{'} 46\farcs 84 \pm 0\farcs 1$ & $7\farcs1$ &
          25.4$\pm0.1$ & 24.7$\pm0.3$\tablenotemark{e}
        & 20.4$\pm 0.1$ & 18.4$\pm 0.1$ & 18.4$\pm 0.2\tablenotemark{d}$ \\
\enddata
\tablenotetext{a}{STIS 50CCD magnitudes.}
\tablenotetext{b}{Magnitude from \citet{Kaplan01}.}
\tablenotetext{c}{Magnitude based on WFPC2 F675W image \citet{Kaplan01}.}
\tablenotetext{d}{Magnitude based on measurements of K band image of Cas A from \citet{GF01}. }
\tablenotetext{e}{Magnitude from Ryan et al.\ (2001).}
\end{deluxetable}

\begin{deluxetable}{llrrrcl}
\tabletypesize{\scriptsize}
\tablecaption{Observed Optical and Near IR Magnitudes for Radio-Quiet Neutron Stars}
\tablewidth{0pt}
\tablehead{
\colhead{Object} & \colhead{SNR} &  \colhead{$R$} &
\colhead{$J$} &
\colhead{$H$} &
\colhead{$K_{s}$} &  \colhead{Refs.} \\
\colhead{} & \colhead{} &
\colhead{(mag)} &
\colhead{(mag)} &
\colhead{(mag)} &
\colhead{(mag)}  & \colhead{} } 
\startdata
{\underline{CCOs}}  \\
Cas A XPS & G111.7$-$2.1 (Cas A) & $\geq$27.8  & $\geq$26.2$\tablenotemark{a}$  & $\geq$24.6$\tablenotemark{a}$ &    $\geq$21.2  &  1--4 \\
CXOU J0852$-$4617       & G266.2$-$1.2 (Vela Junior)            & $\geq$22.0  &  \nodata
   &  \nodata        & \nodata     &  5, 6 \\
1WGA J1713$-$3949     & G347.3$-$0.5       &  \nodata    & \nodata     & \nodata
        & \nodata     &  7 \\
RX J0822$-$4300       & G260.4$-$3.4 (Puppis A)    & $\geq$24.8  &  \nodata    &  \nodata    & \nodata         &  8, 9 \\
1E 1207$-$5209\tablenotemark{b} & G296.5+10.0 (PKS 1209$-$51/52)& $\geq$26.6  & $\geq$23.5  & $\geq$22.4        & $\geq$22.0  &  10 \\
RX J1853+0040         & G33.6+0.1 (Kes 79)       & $\geq$24.9  &  \nodata    &  \nodata        & \nodata     &  11, 12  \\
{\underline{AXPs}}    &                          &             &             &                           &             & \\
XTE J1810$-$197       & \nodata                  &  $\geq$21.5    &
$\geq$23.0   & 22.0--22.7\tablenotemark{c}      & 20.8--21.4\tablenotemark{c}  & 13--16 \\
1E 1048$-$5937        & \nodata                  &  $\geq$24.8    & 21.7--23.4\tablenotemark{c}
 &20.8--($\geq$21.5)\tablenotemark{c}     & 19.4--21.3\tablenotemark{c}  & 17--20 \\
1E 1841$-$045
& G27.4+0.0 (Kes73)        &
$\geq$23.0  &
$\geq$22.1  &  $\geq$20.7   &  $\geq$19.9 & 21--23 \\
CXOU J0100$-$7211\tablenotemark{d} & \nodata &  
24.2\tablenotemark{e}   & \nodata  & \nodata        & \nodata     & 24--27 \\
1RXS J1708$-$4009\tablenotemark{d}     & \nodata                  &  $\geq$26.5    &
   20.9      & 18.6;18.85\tablenotemark{f}   & 18.3;17.5\tablenotemark{f}  & 28 \\
4U 0142+614           & \nodata                  &   24.9      &   \nodata   &  \nodata        & 19.7--20.2\tablenotemark{c}  & 29,30 \\
1E 2259+586           & G109.1$-$1.0 (CTB 109)   &  $\geq$26.4    &    $\geq$23.8  &
  \nodata       & 20.4--21.7\tablenotemark{c}  & 31--34 \\
\enddata
\tablenotetext{a}{$m_{\rm F110W} \approx J$; $m_{\rm F160W} \approx H$.}
\tablenotetext{b}{\citet{Moody05} detected an M dwarf in the {\sl Chandra} 
                  error circle, with $J=21.7$, $H=21.2$, and $K_s=20.7$. 
                  The magnitudes in table are $3\sigma$ detection limits from that observation.}
\tablenotetext{c}{Source variable.}
\tablenotetext{d}{Proposed counterpart unconfirmed}.
\tablenotetext{e}{$m_{\rm F606W}$ based on {\sl HST} WFPC2 observations.}
\tablenotetext{f}{\citet{Israel04} quote different magnitudes
from NTT ($H$ and $K_{\rm s}$) and CFHT ($H$ and $K'$) data.}
\tablenotetext{~}{References -- 
1: \citet{Kaplan01}, 2: \cite{Ryan01} 3: \citet{Fes02}, 4: this work,
5: \citet{Pavlov01}, 6: \citet{Pell02},
7: \citet{cassam04}
8: \citet{Petre96}, 9: \citet{Wang_Chak02a},
10: \citet{Moody05},  
11: \citet{Seward03}, 12: \citet{Gotthelf05}, 
13: \citet{Israel04}, 14: \citet{Gotthelf04a}, 15: \citet{Ibrahim04}, 16: \citet{Rea04},
17: \citet{Wang_Chak02b}, 18: \citet{Israel02},  19: \citet{Durant04},  20: \cite{Durant05a}
21: \citet{Mere01}, 22: \citet{Wachter04}, 23: \citet{Durant05c}
24: \citet{Lamb02}, 25: \citet{Naze03}, 26: \citet{McGarry05}, 27: \citet{Durant05b} 
28: \citet{Israel03},
29: \citet{Hulleman00}, 30: \citet{Hulleman04}, 
31: \citet{Hulleman01},  32:  \citet{Kaspi03}, 33: \citet{Israel03}, 34: \citet{Tam04}
 }
\end{deluxetable}

\begin{deluxetable}{lrccccccccc}
\tabletypesize{\scriptsize}
\tablecaption{Observed X-Ray and NIR Fluxes and Estimated Luminosities 
              for Radio-Quiet Neutron Stars}
\tablewidth{0pt}
\tablehead{
\colhead{Object} & \colhead{$P$} &   \colhead{Age\tablenotemark{a}}   & \colhead{$D$} & \colhead{$N_{\rm H,22}$} &
\colhead{$F^{\rm obs}_{\rm X,-12}$} &\colhead{$F^{\rm obs}_{\rm H,-16}$} 
& \colhead{$F^{\rm obs}_{\rm Ks,-16}$} &
\colhead{$L_{\rm X,33}$} & \colhead{$L_{\rm H,30}$} & \colhead{$L_{\rm Ks,30}$}  \\
\colhead{} &  \colhead{(s)} &  \colhead{(kyr)}  & \colhead{(kpc)} & 
& & & & }
\startdata
{\underline{CCOs}} &          &       &         &        &       & &  &               &         &             \\
Cas A XPS      & \nodata  &  0.3  & 3.4     &  1.2   & 0.8 ~ &$\leq$0.7\tablenotemark{b} &  $\leq$5 ~     &  0.75  & $\leq$0.3     & $\leq$1.3      \\
CXOU J0852$-$4617    & \nodata  &  1--3 &   2     &  0.4   & 1.4 ~ & \nodata  &  \nodata
  &  0.29  & \nodata     & \nodata     \\
1WGA J1713$-$3949  & \nodata  & 1--3 &   2    &  0.5   & 2.9 ~ & \nodata  &  \nodata      &   0.62  & \nodata & \nodata \\
RX J0822$-$4300    & \nodata  &  3--4 &   2     &  0.3   & 4.5 ~ & \nodata  &  \nodata
  &  0.33  & \nodata     & \nodata     \\
1E 1207$-$5209     &   0.4    &  3--20 &   2     &  0.1  & 1.9 ~& $\leq$5   &  $\leq$2 ~
  &  0.084  & $\leq$0.28     & $\leq$0.11     \\
RX J1853+0040      &   0.1    &  5--8  &   7     &  1.5   & 0.2 ~ & \nodata  &  \nodata      &  1.1  & \nodata     & \nodata     \\
{\underline{AXPs}} &          &       &         &        &         &               &         &             \\
XTE J1810$-$197      &   5.5    & [2.4] & 5      &  1.1   & 0.5--90  & 4--8  &  4--7  
   &  0.2--65 & 3--6 & 2--4        \\
1E 1048$-$5937     &  6.5     & [4.3] & 3       &  1.0   & 7--30 ~& $(\leq$12)--24 &  4--25     &  5--25 & $(\leq$3)--6  &  0.8--4     \\
1E 1841$-$045      &  11.8 &[4.5] & 7    &  2.5   &  20 ~ & $\leq$26 & $\leq$15          &   110 & $\leq$140  &   $\leq$35        \\
CXOU J0100$-$72111 &  8.0 & [6.8] & 57      &  0.3   & 1.9  ~& \nodata  &  
\nodata      &   39  & \nodata  & \nodata     \\
1RXS J1708$-$4009\tablenotemark{c}  &  11.0    & [9.0] & 8       &  1.4   & 40 ~ & 180   &   68         &  490 & 460   &  110        \\
4U 0142+614        &  8.7     & [70]  & 3       &  0.9  & 120 ~ & \nodata   &  12--19     &  89 & \nodata    &  2--3       \\
1E 2259+586        &   7.0 &[220] & 3       &  1.0   & 30--90~ & \nodata &  3--10     &  20--60 & \nodata &  0.6--2     \\
\enddata
\tablecomments{CCO and AXP values listed for period, age, and distances (cols.\ 2--4) are taken 
from \citet{Pavlov02a,Pavlov04} and \citet{Woods04},
with updates from the references listed in Table 5.
Hydrogen column densities (col. 5) are in units of $10^{22}$ cm$^{-2}$. 
Observed (absorbed) X-ray flux for the 0.5--10 keV range (col. 6) and H and K$_{s}$ band fluxes (cols.\ 7--8)
are in units of  $10^{-12}$ erg cm$^{-2}$ s$^{-1}$ and $10^{-16}$ erg cm$^{-2}$ s$^{-1}$,
respectively. Adopting the distances listed in column 4,  
estimated X-ray luminosities in the 2--10 keV range in units of $10^{33}$ erg s$^{-1}$
and H and K$_{s}$ band luminosities in units of $10^{30}$ erg s$^{-1}$
are given in columns 9--11.}

\tablenotetext{a}{Ages for the CCOs are associated SNR age estimates.
Age estimates for the AXPs (listed in brackets) are 
spin-down values 
taken from \citet{Woods04}. For J1810$-$197
and J0100$-$7211, we list estimates from \citet{Gotthelf04a}
and \citet{McGarry05}, respectively. Age estimates for
the two AXPs thought associated with SNRs, 1841$-$045 in G24.7+0.0 and
2259$+586$ in G109.1$-$1.0, are considerably larger than the SNR ages,
namely, 0.5--2 kyr \citep{Helfand94} and 8--10 kyr \citep{Sasaki04},
respectively. }
\tablenotetext{b}{F160W $\approx$ H.}
\tablenotetext{c}{Proposed NIR counterpart unconfirmed.}
\end{deluxetable}

\newpage

\begin{figure}
\plotone{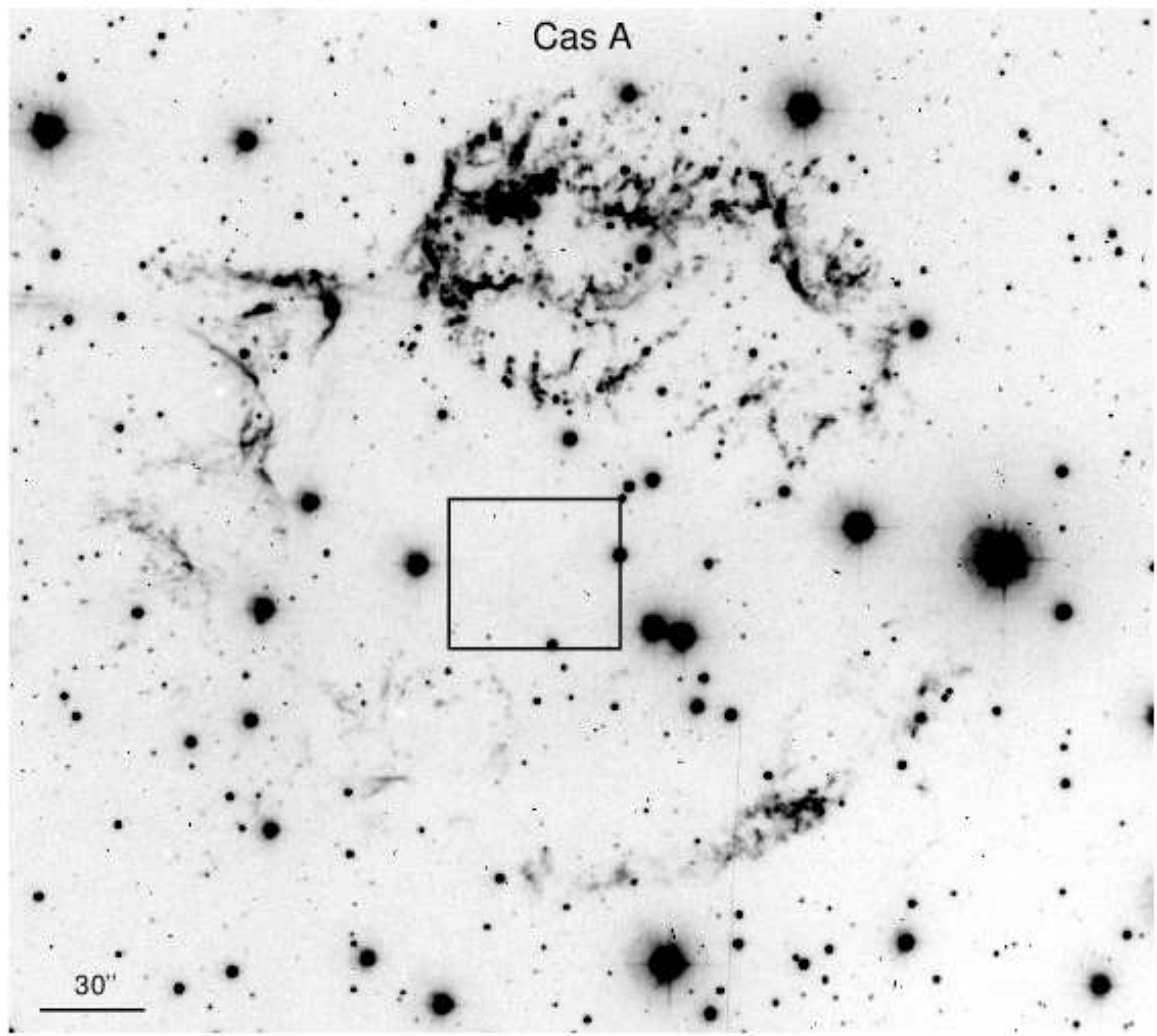}
\caption{MDM R band image of Cas A with the field of view of the {\sl HST} STIS image
of the remnant's central region (shown in Fig.\ 2) marked. }
\label{fig:MDMimage}
\end{figure}

\begin{figure}
\epsscale{0.80}
\plotone{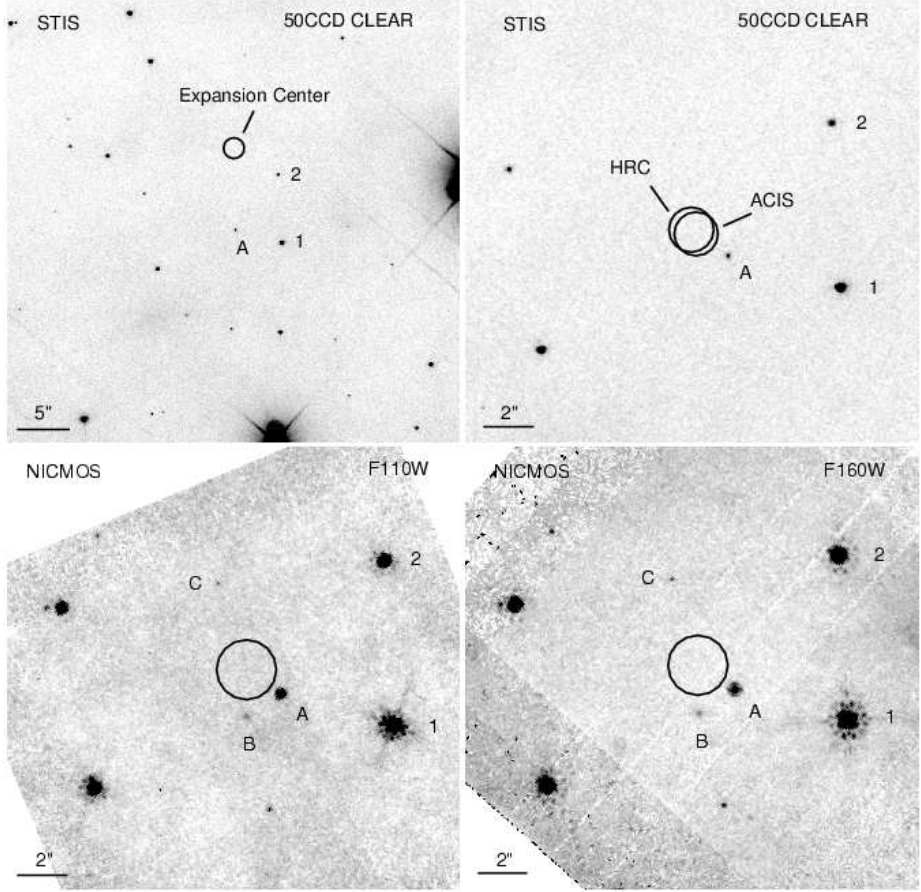}
\caption{ {\sl HST} images of the Cas A central region near the X-ray point source (XPS). 
Upper left panel shows the 2001 STIS
image of the remnant's central region with the \citet{Thor01} expansion center indicated 
($\alpha = 23^{\rm h}23^{\rm m}27 \fs 77\pm 0\fs 05$,
$\delta = 58^{\rm o}48'49 \farcs 4 \pm 0\farcs 4$ ). 
Upper right panel shows an enlargement of the STIS image with the 95\% confidence level circles 
marked for  {\sl Chandra} ACIS-S (radius = $0 \farcs 9$) and HRC-S (radius = $0 \farcs 9$)
centered on the respective mean positions for each data set as listed in Table 2.
Star `A' lies closest to the nominal XPS position but is likely a foreground late-type star \citep{Kaplan01}.
Lower panels show NICMOS J and H band images of the Cas A central region with
99\% confidence level circles (radius = $1 \farcs 2$) shown centered on our adopted position
for the XPS ($\alpha = 23^{\rm h} 23^{\rm m} 27 \fs 943$,
$\delta = 58^{\rm o} 48' 42 \farcs 51$; see text and Table 2). 
While the NICMOS H band image revealed a few faint, additional 
sources (B and C) near the XPS, no source is detected 
within the {\sl Chandra} error circle. }
\label{fig:HST}
\end{figure}

\begin{figure}
\vskip -7cm
\plotone{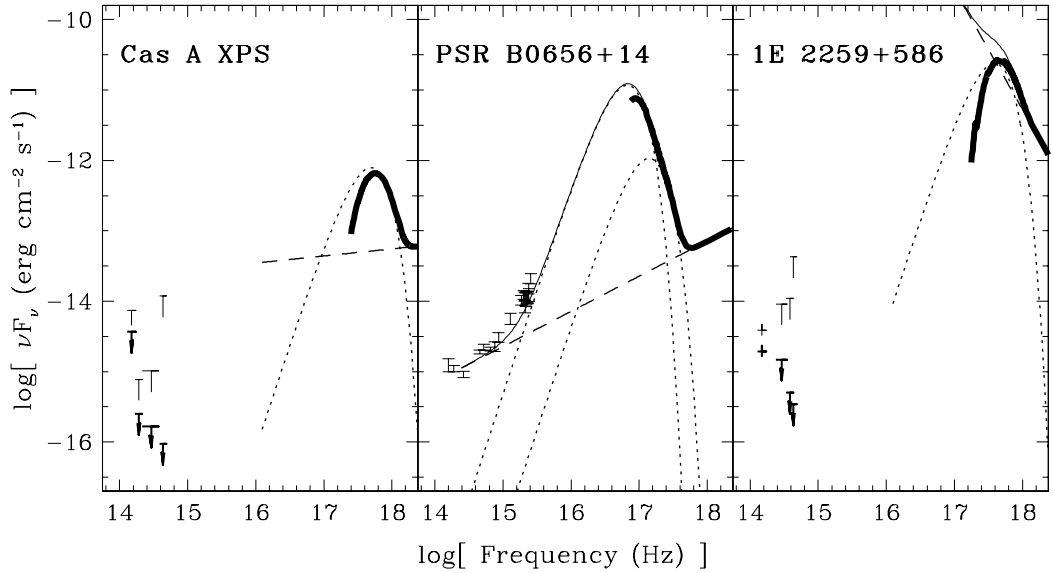}
\caption{Comparisons of the broadband energy spectra of the 
AXP 1E~2259+586 in CTB 109,
PSR B0656+14, and the Cas A XPS. 
Thick and thin solid curves show the observed
and extinction-corrected spectra. Dotted and dashed
lines show the extinction-corrected thermal and power-law 
components of the spectral
fits, respectively (there are two thermal components in the spectrum of
B0656+14). Thick and thin symbols in the NIR-optical domain correspond
to the observed and extinction-corrected detections or upper limits
(only the extinction-corrected points are shown for PSR B0656+14).
For the variable 1E~2259+586,
the X-ray spectral flux is taken
from archival {\sl Chandra} data (observed on 2000 January 11;
see \citealt{Patel01})
while the NIR points are from a Keck I observation
of 1999 June 23/24 \citep{Hulleman01}. 
The AXP was supposedly in a quiescent state during both the NIR and X-ray
observations.  
  }
\label{fig:SEDs}
\end{figure}

\end{document}